\documentclass[5p]{elsarticle}
\usepackage{lineno}
\usepackage{hyperref}
\usepackage{graphicx}
\usepackage{amsmath}
\usepackage{upgreek}
\usepackage{physics}
\usepackage[ulem=normalem]{changes}
\usepackage{amssymb}

\begin{document}
\journal{Nucl. Instrum. Methods Phys. Res. A}
\title{Effects of Magnetic Fields on HPGe Tracking Detectors}
\author{I.~Y.~Lee and A.~O.~Macchiavelli}

\address{Nuclear Science Division, Lawrence Berkeley National Laboratory, 1 Cyclotron Rd, Berkeley, CA 94720, USA}
 
\begin{abstract}
We present a study of magnetic fields effects  on the position resolution and energy response of hyper-pure germanium detectors.  Our results provide realistic estimates of the potential impact on the resolving power of tracking-arrays from (fringe) magnetic fields present when operating together with large spectrometers. 

\noindent
By solving the equations of motion for the electron and holes in the presence of both electric and magnetic fields, we analyzed the drift trajectories of the charge carriers to determine the deviations in the positions at the end point of the trajectories, as well as changes in drift lengths affecting the energy resolution and peak shift due to trapping.  Our results show that the major effect is in the deviation of the transverse (to the electric field direction) position and suggest
that, if no corrective action is taken in the pulse-shape and tracking data analysis procedures, a field strength $\gtrsim$  0.1 T will start to impact the intrinsic position resolution of 2 mm (RMS). At fields  above $\sim$1 T, the degradation of the energy response becomes observable. 
\end{abstract}
\begin{keyword}
Radiation detection \sep High purity germanium detectors \sep $\gamma$-ray tracking and imaging \sep GRETINA/GRETA, AGATA. Magnetic field.
\end{keyword}

\maketitle

\section{Introduction}
\label{sec:intr} 

The structure of nuclei far from the stability line is a central theme of research in nuclear
physics. Key to this program has been the worldwide development of radioactive beam
facilities and novel detector systems, which provide the tools needed to produce and
study these exotic nuclei. In particular, gamma-ray spectroscopy plays a vital and ubiquitous role in these studies~\cite{IY1,JB}.

The $\gamma$-ray tracking technique~\cite{IY2,MAD} marks a major advance in the development of  $\gamma$-ray detector
systems and can provide order-of-magnitude gains in sensitivity compared to existing
arrays. It uses highly-segmented hyper-pure germanium (HPGe) crystals together with advanced
signal processing techniques to determine the location and energy of individual $\gamma$-ray
interactions, which are then combined to reconstruct the incident $\gamma$-ray in a process called
tracking.
A 4$\pi$ tracking-array will be a powerful instrument needed in a broad range of experiments
addressing the intellectual challenges of low-energy nuclear science~\cite{NSAC,NUPECC}. Developments of
these instruments are underway~\cite{IY3} both in the US (GRETINA/GRETA)~\cite{SP,DW,FDR} and Europe~\cite{SA,AGATA}
(AGATA).

In many applications, some of the detector modules are located close to large spectrometers where the magnetic fringe field may not be negligible,
in some cases up to $\approx 0.1$ T. 
For large crystal volumes, the deflection of the charge carriers in the magnetic
field may result in deviations of the derived  interaction point positions, a larger rise times of the signals due to longer drift paths and/or enhanced trapping and de-trapping, and changes in their 
performance are anticipated.  Thus, it is of  importance to understand and predict the performance of HPGe detectors operating in a magnetic field,  particularly in terms of their energy and position resolution, directly affecting the resolving power of the array.
In previous experimental works\footnote{In a relevant work~\cite{AC}, the performance of silicon drift detectors in a magnetic field was studied.}, the response of large coaxial detectors as a function of magnetic field was investigated to assess the impact in nuclear
structure physics~\cite{ASL,KZ,MA} and nuclear medical imaging~\cite{LJH} applications. 
Here we focus on a theoretical study of these effects, using the available empirical data in the references above to benchmark our results.

\section{Calculation}
\label{sec:calc}

\begin{figure}
\centering\includegraphics[trim=140 140 100 100,clip,width=5cm,angle=90]{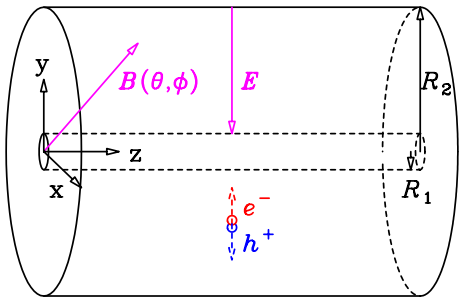}
\caption{\label{fig0} (Color online) Schematic illustration of the coaxial detector, the adopted axes system, and the $\vb*{E}$ and $\vb*{B}$ fields under which the electrons ($e^-$) and holes ($h^+$) drift in the Ge volume.   $\vb*{E}$ is in the radial direction, while the orientation of $\vb*{B}$ is defined  by the angles $\left( \theta,\phi \right)$ with respect to the (x,y,z) reference frame.} 
\end{figure}

The calculations were carried out assuming a coaxial detector with inner radius $R_1$ and outer radius $R_2$. The electric field $\vb*{E}$ is in the radial direction and the magnetic field $\vb*{B}$ could be in any direction defined by the polar angles $\left( \theta,\phi \right)$ relative to the detector axis (z-axis), as schematically shown in Fig.~\ref{fig0}. The trajectories $\vb*{r}(t)$ of electrons and holes follow the drift equation:
\begin{equation}
\frac {d \vb*{r}(t)}{dt} = q \left( \mu \vb*{E} + \mu_H \frac {d\vb*{r}(t)} {dt} \times \vb*{B} \right)
\label{eq1}
\end{equation}
where $\mu$ is the mobility and $\mu_H$ is the Hall mobility. From a common initial position, the separated trajectories of the electron and the hole were solved numerically using their proper values of charge $q$ and mobilities~\cite{IY4}. It is assumed that the electric field direction causes the holes to drift to the outer electrode and the electrons to the inner, as it happens in a typical n-type Ge detector.
From the trajectories, a number of physical quantities can be derived. The key parameters, central to the understanding of the magnetic field effects on a tracking detector, are the position deviations and the change in drift times. The distance between the end positions of the trajectory with and without the magnetic filed gives the position deviation, and the difference in the drift time is similarly obtained. The pulse shape, resulting from the combined contributions from electrons and holes, is calculated using the weighting potential $V(r)$ and the positions determined from their respective trajectories~\cite{Shockley, Ramo}. For the outer electrode we have:
\begin{equation}
V(r) =\frac { \ln(\frac{r}{R_1})} {\ln(\frac{R_2}{R_1})}
\end{equation}
Typical values of the input parameters we used are based on the average properties of GRETA detectors~\cite{SP} and are given in Table~\ref{tab1}.

To estimate the effects on the energy resolution and the shift of the full energy peak, charge trapping distances were introduced for both electrons and holes ($ \lambda_e$ and $\lambda_h$).  The charge attenuation is modeled after Refs.~\cite{TR,MD}:
\begin{equation}
q_e = q_{0e} exp\left( -\frac{\Delta r_e}{\lambda_e}  \right)
\label{eq3}
\end{equation}
and   
\begin{equation}
q_h = q_{0h} exp\left( -\frac{\Delta r_h}{\lambda_h} \right)
\label{eq4}
\end{equation}
As the length $\Delta r$ of the trajectory increase with the magnetic field, charge loss due to trapping increases and cause larger peak shift and worse energy resolution.

\begin{table}
  \caption{\label{tab1} Input parameters used in our calculations}
  \bigskip
  \begin{center}
 \begin{tabular}{c c}
    \hline
    \hline
    Parameter&Value \\
    \hline
   $R_1$& 0.5 cm \\
   $R_2$& 4 cm \\
   \hline
   $E$& 1.46 kV/cm \\
   \hline
   $\mu_e$ &0.77 m$^2$/V/s\\
   $\mu_h$ & 0.62 m$^2$/V/s\\
   \hline
   $\mu_{He}$& $3\pi/8 \mu_e$ \\
   
   $\mu_{Hh}$& $3\pi/8 \mu_h$\\
  \hline
  \hline
  \end{tabular}
  \end{center}
\end{table}


\section{Results and Discussions}
\label{sec:resu}

Under the inference of both electric and magnetic fields and following from the solution of Eq.~\ref{eq1}, a typical drift trajectory of a charge carrier in a HPGe detector is not a straight line. We now present some examples of our calculations  at a magnetic field of 1 T.  

\begin{figure}
\centering\includegraphics[trim=40 100 40 100,clip,width=7cm,angle=90]{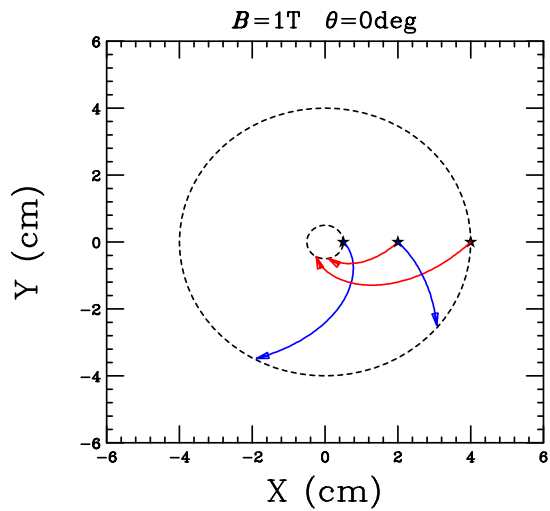}
\caption{\label{fig1} (Color online) Example of $e^-$ (red) and $h^+$ (blue) drift trajectories calculated for a 1 T longitudinal magnetic field. Three initial interaction points (stars) are shown.} 
\end{figure}

Figure~\ref{fig1} shows the electron (red) and hole (blue) trajectories, corresponding to three initial positions.
In this case, the magnetic field is parallel to the z-axis, $\theta=0\deg$, and thus perpendicular to the radial electric field. The trajectories, having spiral shapes, are confined in the x-y plane .

\begin{figure}
\centering\includegraphics[trim=40 100 40 100,clip,width=7cm,angle=90]{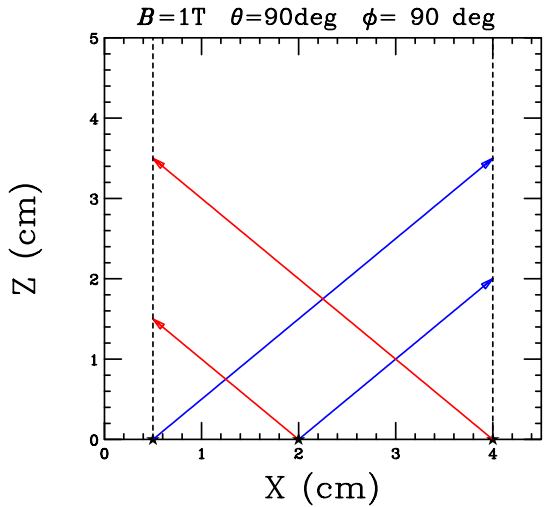}
\caption{\label{fig2} (Color online) Similar to Fig.~\ref{fig1} but for a 1 T magnetic field along the y-axis, with the three initial interaction points (stars) indicated. In these cases the carriers drift in the xz plane.} 
\end{figure}

The example in Figure ~\ref{fig2} corresponds to the case when the magnetic field is in the y-direction, perpendicular to both the z- and x-axis ($\theta=90\deg, \phi=90\deg)$. The position vector of the starting point of the charge carrier is on the x-direction. The trajectories are in the x-z plane and this is the only situation for a coaxial detector where the they are straight lines. In general, given the positioning of the array with respect to the spectrometer, the magnetic field will have an arbitrary direction $\left( \theta,\phi \right)$ relative to the axis of a particular detector and the trajectories will be more complicated.

\begin{figure}
\centering\includegraphics[trim=40 100 40 100,clip,width=7cm,angle=90]{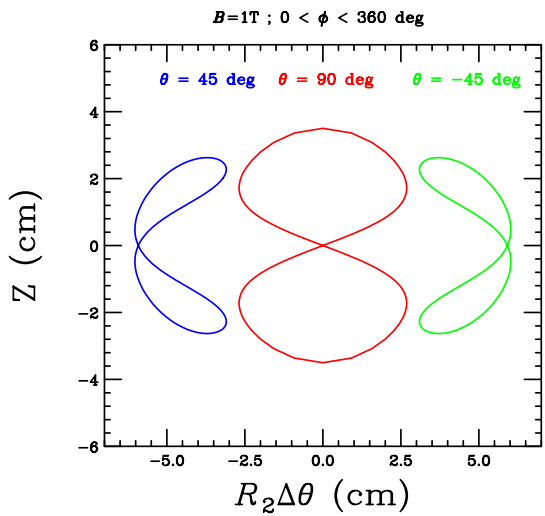}
\caption{\label{fig3} (Color online) Examples of the curves traced by the end points of the $h^+$ trajectories, projected on the cylindrical outer surface of the detector. The three conditions are indicated in the plot. At zero field the end point is at $\left( 0,0 \right)$.} 
\end{figure}

\subsection{Position deviations}

\begin{figure}
\centering\includegraphics[trim=40 100 40 100,clip,width=7cm,angle=90]{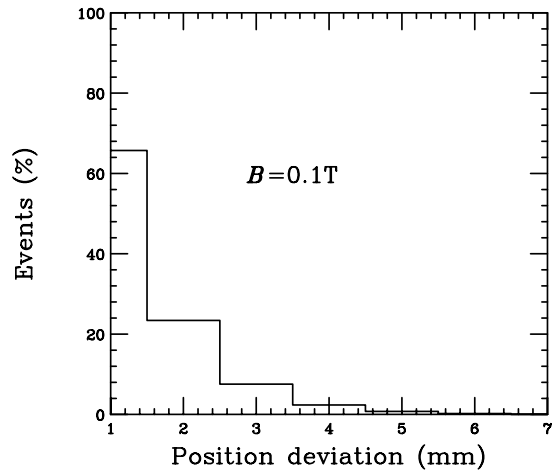}
\caption{\label{fig4} Calculated distribution of the position deviations for $h^+$ at  a 0.1 T field.  } 
\end{figure}

Having calculated the carriers trajectories, we can now analyze the distribution of the end point positions.
In Figure~\ref{fig3} we show the example of a hole trajectory on the detector outer surface, for various three magnetic field directions. The end points trace a 2-dimensional curve in the axial (z) and tangential (x-y) direction. The deviation of the end point position under magnetic field from that without magnetic field causes a two-dimensional position error in the tracking detector.  Figure ~\ref{fig4} shows the distribution of position deviations on the outer surface for the hole trajectories for all the magnetic field directions and all starting positions with a magnetic field strength of 0.1 T. In this case the deviation averaged over all events is $\approx$ 1 mm, and about 65\% of the deviation is less than 1 mm. The  average position deviation increases linearly with the magnetic field strength as seen in Table~\ref{tab2}.

\begin{table}
  \caption{\label{tab2} Average position deviation and increase in drift time as functions of $B$}
  \bigskip
  \begin{center}
 \begin{tabular}{c c c}
 \hline
    \hline
   $B$ (T) &  Deviation (mm) &  Drif time increase (\%) \\
    \hline
  0 & 0& 0  \\
   0.1 & 0.9& 0.3  \\
    0.2 & 1.8& 1.9  \\
    0.3 & 2.8&3.5\\
  \hline
  \hline
  \end{tabular}
  \end{center}
\end{table}

\begin{figure}
\centering\includegraphics[trim=40 100 40 100,clip,width=7cm,angle=90]{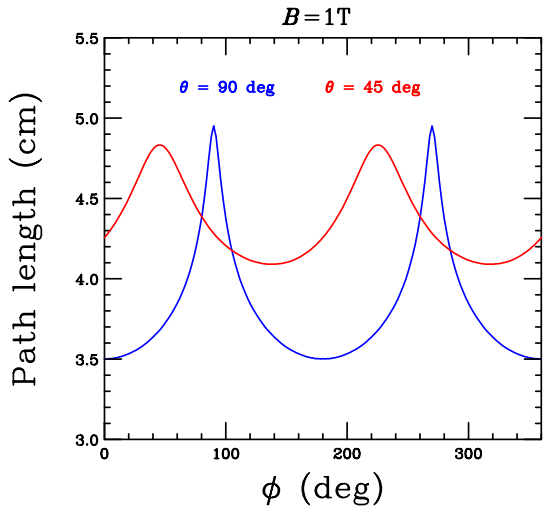}
\caption{\label{fig5} (Color online) Path lengths as a function of the magnetic field azimuthal orientation for two cases of the polar angle as indicated. } 
\end{figure}

In addition to the two-dimensional positional sensitivity from the segmentation, tracking detectors use the drift time information for position determination in the drift direction. As the magnetic field increases the drift path length this introduces a position error in the drift direction. The example in Figure ~\ref{fig5} shows the drift path length as functions of the magnetic field directions for $B=1$T. The drift time deviations, averaged over the field orientations,
are also given in Table~\ref{tab2} for several magnetic field strengths. The longer drift time will result in a longer raise time and change the pulse shape.    The effect is relatively small at low field and increases approximately quadratically at high field

\subsection{Energy resolution and peak shift}

As it follows from Eqs.~\ref{eq3} and \ref{eq4}, charge losses due to trapping will increase as the path length increases, resulting in the reduction of the pulse height and thus peak energy shift. In addition, due to the variation of path length depending on the starting position and the relative angle between the drift direction and the field direction, the peak width will increase and degrade the energy resolution. 

\subsection{Comparison with experimental data}

\begin{figure}
\centering\includegraphics[trim=40 100 40 100,clip,width=7cm,angle=90]{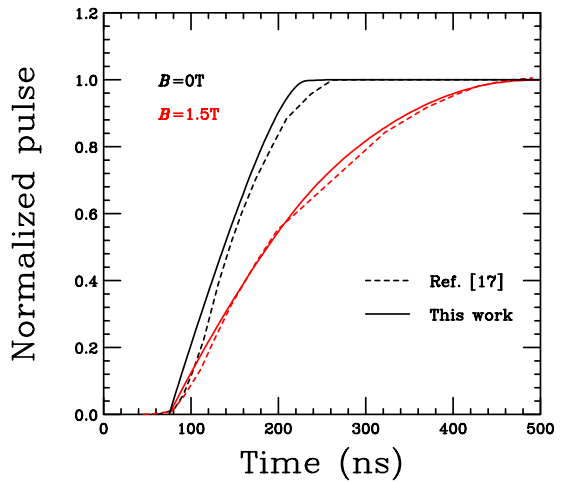}
\caption{\label{fig6} (Color online) Comparison of calculated pulse shapes (solid lines) to the measurements in Ref.~\cite{LJH} (dashed lines) for $B=0$ (black) and $B=1.5$ T (red).  The calculations were normalized to qualitatively reproduce the zero field data. } 
\end{figure}

In order to benchmark our calculations, we contrast our results with selected data from Refs.~\cite{ASL,KZ,LJH}. First we consider a typical example of the effect of the magnetic field on the pulse shape~\cite{LJH}, shown in Fig.~\ref{fig6}. While a fully quantitative comparison may not be possible given that we do not know the full details of the HPGe detector used in that investigation, such as the depletion voltage and bias voltage, it is clear that the effects are well reproduced. The time response of the system was adjusted to reproduce the zero field pulse.

\begin{figure}
\centering\includegraphics[trim=40 100 40 100,clip,width=7cm,angle=90]{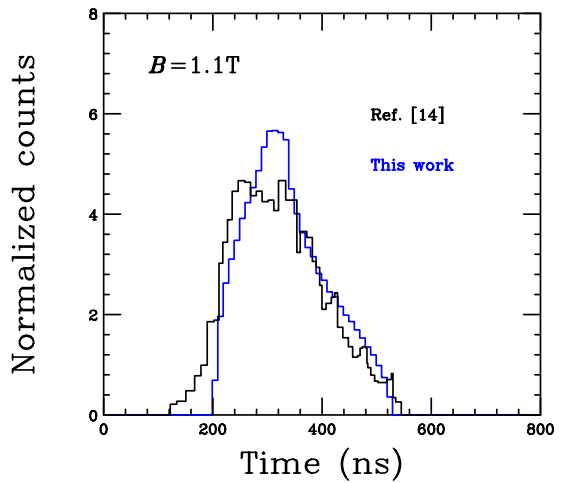}
\caption{\label{fig7} (Color online) Similar to Fig.~\ref{fig6}, comparing 
the calculated rise-time distribution (blue) with the data in Ref.~\cite{ASL} (black). The centroids of the distributions were aligned for display purposes. }

\end{figure}

Similarly, when comparing the measured rise-time distributions in Ref.~\cite{ASL} with our calculations in Fig.~\ref{fig7}, the impact of the magnetic field appears to be captured in the model assumptions. In this case, the distributions were aligned to a common centroid.  

\begin{figure}
\centering\includegraphics[trim=40 100 40 100,clip,width=7cm,angle=90]{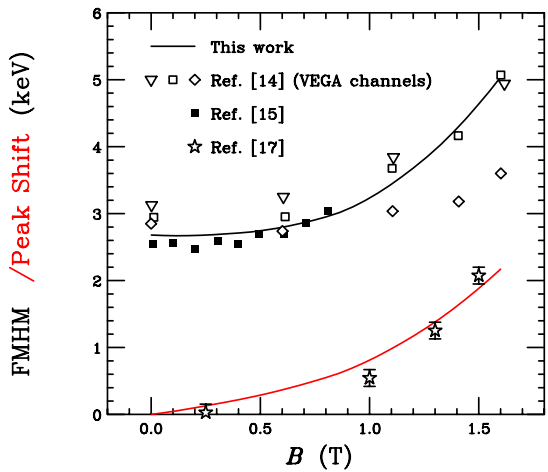}
\caption{\label{fig8} (Color online) Energy response, FWHM (black line) and peak shift (red) as a function of $B$ compared to results of Refs.~\cite{ASL,KZ,LJH}. Note that in our calculations the intrinsic FWHM at  $B$=0 was adjusted to agree with the data. } 
\end{figure}

Finally, we show in Fig.~\ref{fig8} the calculated dependence of the peak shift (negative) and the energy resolution (FWHM) for a 1.33 MeV $^{60}$Co line as a function of magnetic field strength together with data from Refs.~\cite{ASL,KZ,LJH}. Here, the initial FWHM was adjusted to be consistent with the measurements. Charge trapping distances (see Eqs.~\ref{eq3} and \ref{eq4}), $\lambda_{e}=6.0$ m and $\lambda_{h}=2.1$ m, were used to reproduce the results from Ref.~\cite{LJH}, while 4.5 m and 1.6 m respectively were used for the results from Refs.~\cite{ASL,KZ}. Once again our results are in good agreement with the empirical observations. 

\section{Conclusions}
In this work we have studied the effects of magnetic field on the position resolution and energy response of HPGe tracking detectors, such as those used in GRETA. We believe our study provides a simple and robust framework to estimate the potential impact of (fringe) magnetic fields on the resolving power of tracking arrays, when operating together with large spectrometers. This situation is likely to be encountered in a large number of experiments running at rare isotope facilities.   

The major effect is in the deviation of the transverse (to the electric field direction) position at the end point of the trajectories due to the deflection of the electron and hole paths. In a field of $\approx$ 0.1 T this deviation, averaged over all magnet field orientations and interaction points positions, is $\approx$ 1mm with 65\% of the cases less than 1 mm. In the longitudinal direction the averaged drift time increases by $\approx$ 0.3\%, which equals to a 0.1 mm deviation. Thus, if no corrective action is taken in the signal-decomposition and tracking data analysis procedures, to limit the deviation below the intrinsic position resolution of 2 mm (RMS)~\cite{SP,DW} the maximum allowed field strength is $\approx$ 0.1 T. At field strengths higher than 1 T, the degradation of the energy response (resolution and peak shift) becomes observable. This effect is caused by the increase in charge trapping due to the lengthening of the trajectory. 

\section*{Acknowledgments}
This work was supported by the U.S. Department of Energy, Office of Nuclear Physics, under contract no DE-AC02-05CH11231.  AOM would like to thank Prof. J. Men\'endez  (ASU)
for enlightening  discussions.
\bibliography{references}

\begin{thebibliography}{10}
\expandafter\ifx\csname url\endcsname\relax
  \def\url#1{\texttt{#1}}\fi
\expandafter\ifx\csname urlprefix\endcsname\relax\def\urlprefix{URL }\fi
\expandafter\ifx\csname href\endcsname\relax
  \def\href#1#2{#2} \def\path#1{#1}\fi

\bibitem{IY1}
{I. Y. Lee}, {M. A. Deleplanque}, {K. Vetter},
  \href{https://doi.org/10.1088%2F0034-4885%2F66%2F7%2F201}{Developments in
  large gamma-ray detector arrays}, Reports on Progress in Physics 66~(7)
  (2003) 1095--1144.
\newblock \href {https://doi.org/10.1088/0034-4885/66/7/201}
  {\path{doi:10.1088/0034-4885/66/7/201}}.
\newline\urlprefix\url{https://doi.org/10.1088%2F0034-4885%2F66%2F7%2F201}

\bibitem{JB}
J.~Eberth, J.~Simpson,
  \href{http://www.sciencedirect.com/science/article/pii/S0146641007000828}{{{From
  Ge(Li) detectors to gamma-ray tracking arrays - 50 years of gamma
  spectroscopy with germanium detectors}}}, Progress in Particle and Nuclear
  Physics 60~(2) (2008) 283 -- 337.
\newblock \href {https://doi.org/https://doi.org/10.1016/j.ppnp.2007.09.001}
  {\path{doi:https://doi.org/10.1016/j.ppnp.2007.09.001}}.
\newline\urlprefix\url{http://www.sciencedirect.com/science/article/pii/S0146641007000828}

\bibitem{IY2}
{I.Y. Lee},
  \href{http://www.sciencedirect.com/science/article/pii/S0168900298010936}{Gamma-ray
  tracking detectors}, Nuclear Instruments and Methods in Physics Research
  Section A: Accelerators, Spectrometers, Detectors and Associated Equipment
  422~(1) (1999) 195 -- 200.
\newblock \href {https://doi.org/https://doi.org/10.1016/S0168-9002(98)01093-6}
  {\path{doi:https://doi.org/10.1016/S0168-9002(98)01093-6}}.
\newline\urlprefix\url{http://www.sciencedirect.com/science/article/pii/S0168900298010936}

\bibitem{MAD}
{M.A. Deleplanque}, {I.Y. Lee}, K.~Vetter, {G.J. Schmid}, {F.S. Stephens},
  {R.M. Clark}, {R.M. Diamond}, P.~Fallon, {A.O. Macchiavelli},
  \href{http://www.sciencedirect.com/science/article/pii/S0168900299001874}{{GRETA:
  utilizing new concepts in gamma-ray detection}}, Nuclear Instruments and
  Methods in Physics Research Section A: Accelerators, Spectrometers, Detectors
  and Associated Equipment 430~(2) (1999) 292 -- 310.
\newblock \href {https://doi.org/https://doi.org/10.1016/S0168-9002(99)00187-4}
  {\path{doi:https://doi.org/10.1016/S0168-9002(99)00187-4}}.
\newline\urlprefix\url{http://www.sciencedirect.com/science/article/pii/S0168900299001874}

\bibitem{NSAC}
{DOE NSAC},
  \href{https://science.osti.gov/-/media/np/nsac/pdf/2015LRP/2015_LRPNS_091815.pdf}{{{Reaching
  for the Horizon: The 2015 Long Range Plan for Nuclear Science}}}.
\newline\urlprefix\url{https://science.osti.gov/-/media/np/nsac/pdf/2015LRP/2015_LRPNS_091815.pdf}

\bibitem{NUPECC}
NuPECC, \href{http://www.nupecc.org/lrp2016/Documents/lrp2017.pdf}{{{Long Range
  Plan 2017: Perspectives in Nuclear Physics}}}.
\newline\urlprefix\url{http://www.nupecc.org/lrp2016/Documents/lrp2017.pdf}

\bibitem{IY3}
{I-Yang Lee}, J.~Simpson,
  \href{https://doi.org/10.1080/10619127.2010.506124}{{AGATA} and {GRETA}: The
  {Future} of {Gamma-Ray Spectroscopy}}, Nuclear Physics News 20~(4) (2010)
  23--28.
\newblock \href
  {http://arxiv.org/abs/https://doi.org/10.1080/10619127.2010.506124}
  {\path{arXiv:https://doi.org/10.1080/10619127.2010.506124}}, \href
  {https://doi.org/10.1080/10619127.2010.506124}
  {\path{doi:10.1080/10619127.2010.506124}}.
\newline\urlprefix\url{https://doi.org/10.1080/10619127.2010.506124}

\bibitem{SP}
{S. Paschalis}, {I.Y. Lee}, {A.O. Macchiavelli}, et~al.,
  \href{http://www.sciencedirect.com/science/article/pii/S0168900213000508}{{The
  performance of the Gamma-Ray Energy Tracking In-beam Nuclear Array GRETINA}},
  Nuclear Instruments and Methods in Physics Research Section A: Accelerators,
  Spectrometers, Detectors and Associated Equipment 709 (2013) 44 -- 55.
\newblock \href {https://doi.org/https://doi.org/10.1016/j.nima.2013.01.009}
  {\path{doi:https://doi.org/10.1016/j.nima.2013.01.009}}.
\newline\urlprefix\url{http://www.sciencedirect.com/science/article/pii/S0168900213000508}

\bibitem{DW}
D.~Weisshaar, D.~Bazin, P.~Bender, C.~Campbell, et~al.,
  \href{http://www.sciencedirect.com/science/article/pii/S0168900216312402}{{The
  performance of the gamma-ray tracking array GRETINA for gamma-ray
  spectroscopy with fast beams of rare isotopes}}, Nuclear Instruments and
  Methods in Physics Research Section A: Accelerators, Spectrometers, Detectors
  and Associated Equipment 847 (2017) 187 -- 198.
\newblock \href {https://doi.org/https://doi.org/10.1016/j.nima.2016.12.001}
  {\path{doi:https://doi.org/10.1016/j.nima.2016.12.001}}.
\newline\urlprefix\url{http://www.sciencedirect.com/science/article/pii/S0168900216312402}

\bibitem{FDR}
\href{https://sites.google.com/a/lbl.gov/greta/}{{GRETA: The Gamma-Ray Energy
  Tracking Array}}.
\newline\urlprefix\url{https://sites.google.com/a/lbl.gov/greta/}

\bibitem{SA}
S.~Akkoyun, A.~Algora, B.~Alikhani, et~al.,
  \href{http://www.sciencedirect.com/science/article/pii/S0168900211021516}{{AGATA:
  Advanced GAmma Tracking Array}}, Nuclear Instruments and Methods in Physics
  Research Section A: Accelerators, Spectrometers, Detectors and Associated
  Equipment 668 (2012) 26 -- 58.
\newblock \href {https://doi.org/https://doi.org/10.1016/j.nima.2011.11.081}
  {\path{doi:https://doi.org/10.1016/j.nima.2011.11.081}}.
\newline\urlprefix\url{http://www.sciencedirect.com/science/article/pii/S0168900211021516}

\bibitem{AGATA}
\href{https://sites.google.com/a/lbl.gov/greta/}{{{AGATA: The Advanced GAmma
  Tracking Array}}}.
\newline\urlprefix\url{https://sites.google.com/a/lbl.gov/greta/}

\bibitem{AC}
A.~Castoldi, E.~Gatti, V.~Manzari, P.~Rehak,
  \href{http://www.sciencedirect.com/science/article/pii/S0168900297009091}{Performance
  of silicon drift detectors in a magnetic field}, Nuclear Instruments and
  Methods in Physics Research Section A: Accelerators, Spectrometers, Detectors
  and Associated Equipment 399~(2) (1997) 227 -- 243.
\newblock \href {https://doi.org/https://doi.org/10.1016/S0168-9002(97)00909-1}
  {\path{doi:https://doi.org/10.1016/S0168-9002(97)00909-1}}.
\newline\urlprefix\url{http://www.sciencedirect.com/science/article/pii/S0168900297009091}

\bibitem{ASL}
A.~{Sanchez Lorente}, P.~Achenbach, M.~Agnello, T.~Bressani, et~al.,
  \href{http://www.sciencedirect.com/science/article/pii/S0168900206024429}{Performance
  of hpge detectors in high magnetic fields}, Nuclear Instruments and Methods
  in Physics Research Section A: Accelerators, Spectrometers, Detectors and
  Associated Equipment 573~(3) (2007) 410 -- 417.
\newblock \href {https://doi.org/https://doi.org/10.1016/j.nima.2006.12.030}
  {\path{doi:https://doi.org/10.1016/j.nima.2006.12.030}}.
\newline\urlprefix\url{http://www.sciencedirect.com/science/article/pii/S0168900206024429}

\bibitem{KZ}
K.~Szymanska, P.~Achenbach, M.~Agnello, E.~Botta, et~al.,
  \href{http://www.sciencedirect.com/science/article/pii/S0168900208005457}{Resolution,
  efficiency and stability of hpge detector operating in a magnetic field at
  various gamma-ray energies}, Nuclear Instruments and Methods in Physics
  Research Section A: Accelerators, Spectrometers, Detectors and Associated
  Equipment 592~(3) (2008) 486 -- 492.
\newblock \href {https://doi.org/https://doi.org/10.1016/j.nima.2008.04.017}
  {\path{doi:https://doi.org/10.1016/j.nima.2008.04.017}}.
\newline\urlprefix\url{http://www.sciencedirect.com/science/article/pii/S0168900208005457}

\bibitem{MA}
M.~Agnello, E.~Botta, T.~Bressani, M.~Bruschi, S.~Bufalino, et~al.,
  \href{http://www.sciencedirect.com/science/article/pii/S0168900209008961}{Study
  of the performance of hpge detectors operating in very high magnetic fields},
  Nuclear Instruments and Methods in Physics Research Section A: Accelerators,
  Spectrometers, Detectors and Associated Equipment 606~(3) (2009) 560 -- 568.
\newblock \href {https://doi.org/https://doi.org/10.1016/j.nima.2009.04.035}
  {\path{doi:https://doi.org/10.1016/j.nima.2009.04.035}}.
\newline\urlprefix\url{http://www.sciencedirect.com/science/article/pii/S0168900209008961}

\bibitem{LJH}
L.~Harkness, A.~Boston, H.~Boston, P.~Cole, et~al.,
  \href{http://www.sciencedirect.com/science/article/pii/S0168900211003408}{An
  investigation of the performance of a coaxial hpge detector operating in a
  magnetic resonance imaging field}, Nuclear Instruments and Methods in Physics
  Research Section A: Accelerators, Spectrometers, Detectors and Associated
  Equipment 638~(1) (2011) 67 -- 73.
\newblock \href {https://doi.org/https://doi.org/10.1016/j.nima.2011.02.034}
  {\path{doi:https://doi.org/10.1016/j.nima.2011.02.034}}.
\newline\urlprefix\url{http://www.sciencedirect.com/science/article/pii/S0168900211003408}

\bibitem{IY4}
{I. Y. Lee}, {Electron and Hole Drift Velocity in Ge}, GRETINA Technical Note
  GRT-6-061112 (2006).

\bibitem{Shockley}
W.~Shockley, Currents to conductors induced by a moving point charge, Journal
  of Applied Physics 9~(10) (1938) 635--636.
\newblock \href {https://doi.org/10.1063/1.1710367}
  {\path{doi:10.1063/1.1710367}}.

\bibitem{Ramo}
S.~Ramo, Currents induced by electron motion, Proceedings of the IRE 27~(9)
  (1939) 584--585.

\bibitem{TR}
T.~W. Raudorf, R.~H. Pehl,
  \href{http://www.sciencedirect.com/science/article/pii/0168900287912253}{Effect
  of charge carrier trapping on germanium coaxial detector line shapes},
  Nuclear Instruments and Methods in Physics Research Section A: Accelerators,
  Spectrometers, Detectors and Associated Equipment 255~(3) (1987) 538 -- 551.
\newblock \href {https://doi.org/https://doi.org/10.1016/0168-9002(87)91225-3}
  {\path{doi:https://doi.org/10.1016/0168-9002(87)91225-3}}.
\newline\urlprefix\url{http://www.sciencedirect.com/science/article/pii/0168900287912253}

\bibitem{MD}
M.~Descovich, {I.Y. Lee}, {P.N. Luke}, {R.M. Clark}, M.~Cromaz, {M.A.
  Deleplanque}, {R.M. Diamond}, P.~Fallon, {A.O. Macchiavelli}, {E.
  Rodriguez-Vieitez}, {F.S. Stephens}, D.~Ward,
  \href{http://www.sciencedirect.com/science/article/pii/S0168900205004985}{Effects
  of neutron damage on the performance of large volume segmented germanium
  detectors}, Nuclear Instruments and Methods in Physics Research Section A:
  Accelerators, Spectrometers, Detectors and Associated Equipment 545~(1)
  (2005) 199 -- 209.
\newblock \href {https://doi.org/https://doi.org/10.1016/j.nima.2005.01.308}
  {\path{doi:https://doi.org/10.1016/j.nima.2005.01.308}}.
\newline\urlprefix\url{http://www.sciencedirect.com/science/article/pii/S0168900205004985}

\end{thebibliography}

\end{document}